\documentstyle[aps,prb,graphicx,twocolumn]{revtex}

\def\be{\begin{equation}}
\def\ee{\end{equation}}
\def\ba{\begin{eqnarray}}
\def\ea{\end{eqnarray}}

\begin{document}
\draft
\title{Quantum-dot lithium in strong-interaction regime: \\ Depolarization of electron spins by magnetic field
}

\author{S. A. Mikhailov\footnote{Present address: Max-Planck Institut f\"ur Festk\"orperforschung, Heisenbergstr. 1, 70569 Stuttgart, Germany. Electronic address: S.Mikhailov@fkf.mpg.de}}
\address{Theoretical Physics II, Institute for Physics, University of Augsburg, 86135 Augsburg, Germany}
\author{N. A. Savostianova}
\address{Moscow Academy for Device Development and Informatics, Stromynka 20, Moscow 107076, Russia } 

\date{\today}
\maketitle

\begin{abstract}
Magnetic field usually leads to a polarization of electron spins. It is shown that in a system of {\em strongly interacting} particles applying magnetic field may lead to an opposite effect -- depolarization of electron spins. Results of the work are based on an exact-diagonalization study of quantum-dot lithium -- a system of three Coulomb interacting two-dimensional electrons in a parabolic confinement potential.
\end{abstract}


It is well known that, applying a magnetic field to a system of charged particles (e.g. electrons) results in the polarization of their spins. Indeed, being placed in a potential box, $N$ electrons occupy the lowest-energy states (two ones with opposite spins per state), to minimize their total kinetic energy. The total spin of the system $S$ will then be small compared to $N/2$. If the system is now placed in a magnetic field $B$, the Zeeman energy competes with the kinetic energy and favours the spin alignment along the field. The larger the magnetic field, the more electron spins flip over, and at sufficiently strong $B$ the whole system appears to be in the fully spin-polarized state with $S=N/2$. 

This picture is valid if the inter-particle interaction is not too strong and can be neglected. In a system of {\em strongly} interacting particles the interaction can polarize the system already in the absence of magnetic field. An example is a (spin-polarized) Wigner crystal, which was shown to be the ground state of a system of strongly interacting electrons on a positive jellium background.\cite{Tanatar89} Electron-electron interaction forces electrons to occupy higher single-particle energy levels (which reduces the Coulomb energy at the cost of the kinetic one), and favours the ground state with a large total spin $S\simeq N/2$.

Now one can ask, what happens with the system of strongly interacting particles, {\em already spin-polarized} by the inter-particle interaction, if to place it in a magnetic field. At first glance one could think that the magnetic field ${\bf B}=(0,0,B)$ will only split the ground state with $S=N/2$ onto sublevels with different total-spin projections $S_z=\pm N/2, \pm(N/2-1),\dots$. It turns out however that there exist situations when the {\em growing} magnetic field {\em reduces} the total spin $S$ of the system (i.e. forces the back spin-flip of some electrons). This unusual effect is a consequence of a competition of the kinetic, Coulomb and Zeeman energies and may occur only in systems of {\em strongly} interacting particles. In this paper we demonstrate the feasibility of this interesting effect by the example of {\em quantum-dot lithium} -- a system of three two-dimensional (2D) electrons in the harmonic oscillator potential. 

Quantum dots\cite{Jacak97} are ideal physical objects for studying effects of electron-electron correlations. In these systems, realizable in modern semiconductor heterostructures, 2D electrons move in the plane $z=0$ in a lateral harmonic oscillator potential $V({\bf r})=m^\star\omega_0^2 r^2/2$, ${\bf r}=(x,y)$. The Coulomb interaction parameter in dots 
\begin{equation}
\lambda=\frac{l_0}{a_B}=\sqrt{\frac{e^2/a_B}{\hbar\omega_0}}\propto \frac{e^2}{\hbar^{3/2}}
\end{equation} 
is defined as the ratio of the oscillator length $l_0=\sqrt{\hbar/m^\star\omega_0}$ to the effective Bohr radius $a_B$ of the host semiconductor (here $m^\star$ is the electron effective mass, and $\omega_0$ the oscillator frequency). The $\lambda$-parameter can be varied in a wide range, as opposed to natural atoms, which allows one to study properties of the dots as a function of the inter-particle interaction strength. Quantum dots were intensively studied in recent years by a variety of experimental and theoretical techniques, see e.g. Refs. \cite{Jacak97,Ashoori96,Kouvenhouven97,Heitmann93,Schuller98,Maksym00} and references therein.

Many-body quantum states of quantum-dot lithium $(N=3)$ are characterized by quantum numbers $L_z$ (projection of the total angular momentum on the $z$ axis), $S$ and $S_z$. In zero magnetic field energies of these states can be written in the form
\be
E_{L_zSS_z}=
\hbar\omega_0
{\cal F}_{LS}(\lambda),
\label{ELS}
\ee
where ${\cal F}_{LS}$ are some functions of the interaction parameter $\lambda$. The functions ${\cal F}_{LS}(\lambda)$ have been calculated in Ref. \cite{Mikhailov02b} using exact-diagonalization technique (convergency of the method was studied in detail in \cite{Mikhailov02b}; the accuracy of results for the energy of levels was shown to be better than $5\times 10^{-4}$\% at $\lambda\le 10$). Levels (\ref{ELS}) are degenerate with respect to $|L_z|\equiv L$ and $S_z$. In finite magnetic fields we calculate the energy levels and many-body wave functions, as functions of two dimensionless variables $\lambda$ and $\Omega_c=\omega_c/\omega_0$, by the formulas
\be
\frac{E_{L_zSS_z}}{\hbar\omega_0}=\frac{\Omega_c}{2}\left(L_z+g^\star\frac{m^\star}{m_e}S_z\right)+\alpha^2{\cal F}_{LS}\left(\frac{\lambda}{\alpha}\right),
\label{ELSB}
\ee
\be
\Psi_{B\neq 0}({\bf r}_1,\dots,{\bf r}_N)=\Psi_{B=0}(\alpha{\bf r}_1,\dots,\alpha{\bf r}_N),
\label{PsiB}
\ee
$\alpha=(1+\Omega_c^2/4)^{1/4}$, which directly follow from the Hamiltonian of the dot (here $\omega_c$ is the cyclotron energy, $m_e$ is the free electron mass, and $g^\star$ is the effective $g$-factor).

Quantum-dot lithium is an atom, in which strong Coulomb interaction does lead to a full spin polarization in the absence of magnetic field.\cite{Egger99a,Mikhailov02b} In the regime of weak Coulomb interaction ($\lambda<\lambda_{\rm crit}=4.343$, Ref. \cite{Mikhailov02b}) its ground state is partly spin-polarized, $(L,S)=(1,1/2)$. At strong interactions $(\lambda>\lambda_{\rm crit})$ the ground state is fully polarized, with $(L,S)=(0,3/2)$. In the {\em weak} interaction regime $\lambda<\lambda_{\rm crit}$ magnetic field dependencies of the ground state properties were studied in a number of papers earlier, see Refs. \cite{Hawrylak93b,Maksym90,Maksym95,Maksym96}. In the present paper we report results on the ground state properties of quantum-dot lithium for a broad range of parameters, including the regime of {\em strong} Coulomb interaction $\lambda>\lambda_{\rm crit}$. 

\begin{figure}
\includegraphics[width=8.2cm]{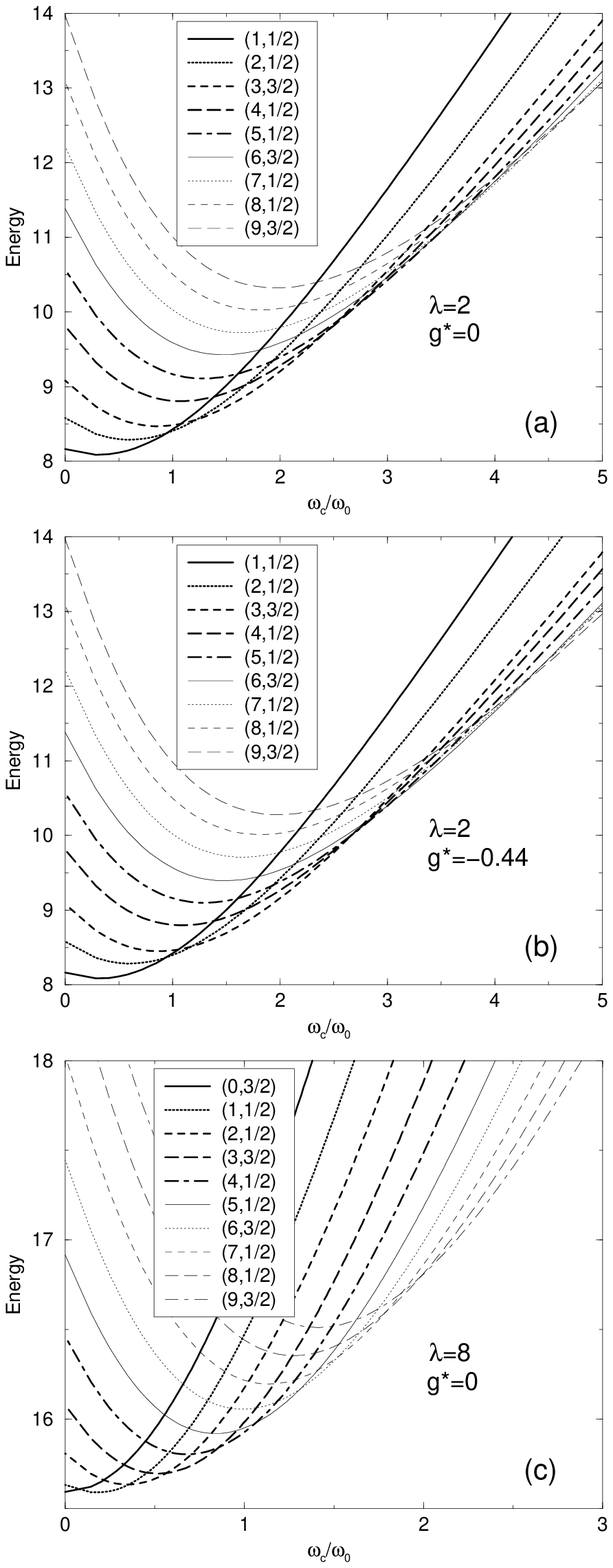}
\caption{Magnetic field dependencies of the ground state levels $E_{L_zSS_z}$ [the curves are labeled as $(-L_z,S_z)$] in the regime of weak (a and b), $\lambda=2$, and strong (c), $\lambda=8$, interactions. Zeeman splitting is ignored in Figures (a) and (c), and included, with parameters of GaAs dots ($g^\star=-0.44$, $m^\star/m_e=0.067$), in Figure (b).
}
\label{energyB}
\end{figure}

\begin{figure}
\includegraphics[width=8.2cm]{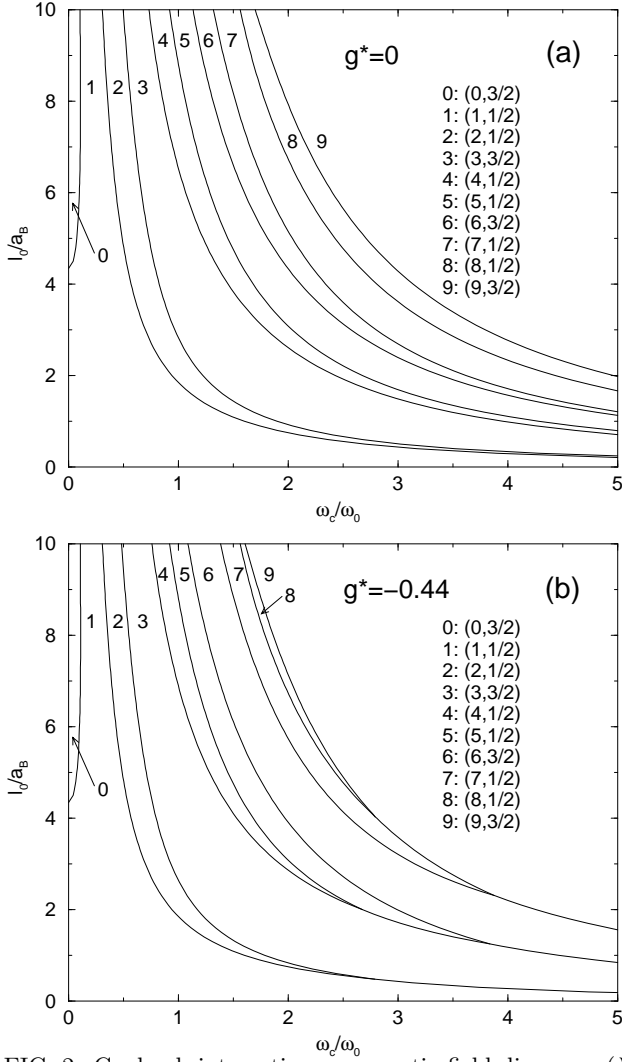}
\caption{Coulomb interaction -- magnetic field diagrams ($\lambda$ vs $\Omega_c$) for the ground states of quantum-dot lithium, without (a) and with (b) Zeeman splitting (GaAs parameters are assumed). The areas 0, 1, 2, ... are labeled as $(-L_z,S_z)$. States with $L\ge 10$ are not shown.
}
\label{lambdaBdiagrams}
\end{figure}

In Figure \ref{energyB} we show energies of a number of ground-state levels as a function of $B$ in the weak and strong interaction regimes (parameters of GaAs quantum dots were used when necessary). When $B$ increases, the system oscillates between partly ($S_z=1/2$) and fully ($S_z=3/2$) spin-polarized ground states, with the total angular momentum of the fully polarized states being multiples of $N$. Between the fully polarized states with $L=3$, 6, 9, $\dots$, pairs of partly polarized states appear, with $L=1$ and 2, 4 and 5, 7 and 8, and so on. If Zeeman splitting is ignored, these partly polarized states can be the ground states at all magnetic fields. If Zeeman splitting is included, only fully spin-polarized ground states with $L=N\times$integer survive at strong $B$. Full $\lambda-\Omega_c$ diagrams for the ground states of quantum-dot lithium with different $L$ and $S_z$ are shown in Figure \ref{lambdaBdiagrams}. Similar results for quantum-dot helium $(N=2)$ were obtained in Ref. \cite{Wagner92} (in helium the system oscillates between the singlet, $S=0$, and triplet, $S=1$, ground states with even and odd values of $L$ respectively). In the weak-interaction regime our results quantitatively agree with those obtained in Ref. \cite{Hawrylak93b} (exact-diagonalization technique) at $\lambda\approx 2$.

A new and interesting feature of the energy spectrum of quantum-dot lithium is seen in Figure \ref{energyB}c (the regime of strong Coulomb interaction, $\lambda=8$) at small magnetic fields. In the strong-interaction regime, $\lambda>\lambda_{\rm crit}$, the $B=0$ ground state is fully spin-polarized and has the total angular momentum $L=0$, while the first excited state is partly polarized and has the total angular momentum $L=1$. When $B$ increases, the ground state $(0,3/2)$ has, mainly, a positive dispersion ($\sim \Omega_c^2$), due to the second term in Eq. (\ref{ELSB}) (the weak negative dispersion due to the Zeeman term $\sim g^\star m^\star \Omega_cS_z$ can be ignored because of the small value of prefactors $g^\star m^\star$). The first excited state $(1,1/2)$ has a strong negative dispersion due to the first term $\sim \Omega_cL_z/2$ ($L_z<0$ in the ground state). These two lowest-energy levels cross each other at a certain critical magnetic field $B=B_{\rm crit}(\lambda)$. The corresponding dimensionless critical parameter $\Omega_c^{\rm crit}(\lambda)$ is of order 0.1 at $\lambda\gtrsim 6$, Figure \ref{lambdaBdiagrams}. Relation between $B_{\rm crit}$ and $\Omega_c^{\rm crit}$, for typical parameters of GaAs quantum dots, can be written in the form 
\be
B_{\rm crit}{\rm (T)}\approx 0.83 \frac{\Omega_c^{\rm crit}(\lambda)}{\lambda^2},
\ee
which gives $B_{\rm crit}\simeq 1.2$ mT at $\lambda\simeq 8$. At $B\gtrsim B_{\rm crit}$ the total spin and its projection are $1/2$ and $+1/2$ in the ground state, while at $B<B_{\rm crit}$ they were $3/2$ and $+3/2$, respectively. The growing magnetic field thus causes a back spin-flip of one electron in the dot. This somewhat unexpected result is a direct consequence of strong Coulomb interaction in the dot at $\lambda>\lambda_{\rm crit}$. In the weak interaction regime $\lambda\le \lambda_{\rm crit}$ the critical field $B_{\rm crit}(\lambda)$ is zero, and the effect is absent. The assumption on the purely parabolic confinement potential in the dot is not essential for this effect. It should also be seen in three-electron systems with non-parabolic confinement, if the Coulomb interaction is sufficiently strong to polarize the system in the absence of magnetic field. The nature of the $L_zSS_z$ states (behaviour of electron and spin densities, as well as pair-correlation functions) can be understood on the basis of Eq. (\ref{PsiB}) and results of Ref. \cite{Mikhailov02b}.

\begin{figure}
\includegraphics[width=8.2cm]{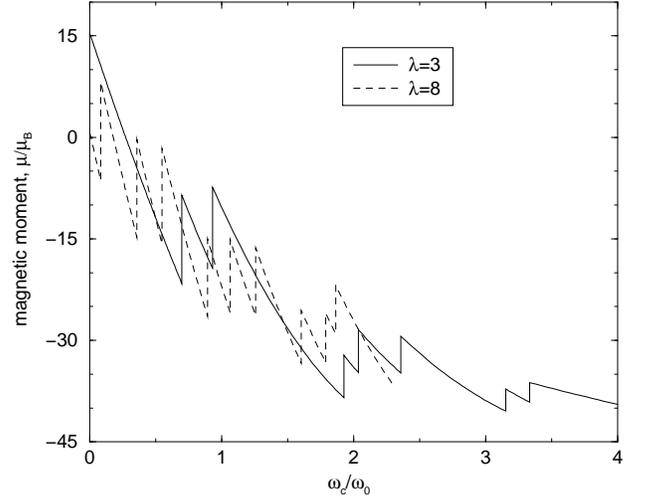}
\caption{Magnetic moment (in the Bohr magneton units) of quantum-dot lithium in the moderate ($\lambda=3$) and strong ($\lambda=8$) Coulomb-interaction regimes. Zeeman splitting is included.
}
\label{moment}
\end{figure}

Finally, we briefly discuss $B$-dependencies of the magnetic moment of the dot, $\mu=-\partial E_{GS}/\partial B$, at zero temperature ($E_{GS}$ is the ground-state energy). Figure \ref{moment} exhibits two typical $\mu(B)$-curves, in the weak and strong Coulomb interaction regimes. The magnetic moment strongly oscillates as a function of $B$, in accordance with the energy curves in Figure \ref{energyB}. At low magnetic fields, the quantum-dot lithium atom is paramagnetic in the regime of weak interactions, and mainly diamagnetic in the regime of strong interactions (except of the region of the very small fields, where it is paramagnetic due to the small $g^\star S_z$ contribution to the energy).

To summarize, we have reported results of an exact-diagonalization study of quantum-dot lithium in finite magnetic field. In the regime of strong Coulomb interaction, which was not treated in the literature so far, we have found a new and unexpected effect of spin depolarization of the quantum-dot lithium atom by weak magnetic field (the total spin $S$ changes from $S=3/2$ to $S=1/2$ with the growth of $B$). The predicted effect is the case in the regime of strong Coulomb interaction $\lambda>\lambda_{\rm crit}=4.343$ and in magnetic fields of order of 1 mT. It could be observed in capacity-, transport-, and Raman-spectroscopy experiments on quantum-dot systems.

\acknowledgments
The work was supported by the Deutsche Forschungsgemeinschaft (SFB 484). We are indebted to Ari Harju for helpful comments.


\end{document}